\documentclass[10pt]{article}
\usepackage{latexsym}
\usepackage{amssymb}
\usepackage{amsmath}
\usepackage{amscd}
\usepackage{amsthm}
\usepackage[left=2.5cm,top=2.5cm,right=2.5cm,bottom=2.5cm]{geometry}

\usepackage[dvips]{graphicx}
\usepackage{hyperref}

\begin{document}
\begin{center}
\large{\bf {Dissipative Future Universe without Big Rip}}\\ 
\vspace{10mm}
\normalsize{Anil Kumar Yadav}\\
\vspace{4mm}
\normalsize{Department of Physics, Anand Engineering
College, Keetham, Agra-282 007, India} \\
\vspace{2mm}
\normalsize{E-mail: abanilyadav@yahoo.co.in, akyadav@imsc.res.in}
\end{center}
\vspace{10mm}
\begin{abstract} 
The present study deals with dissipative future universe without big rip in context of Eckart formalism. 
The generalized chaplygin gas, characterized by equation of state $p=-\frac{A}{\rho^\frac{1}{\alpha}}$, has been 
considered as a model for dark energy due to its dark-energy-like evolution at late time. It is demonstrated that, if 
the cosmic dark energy behaves like a fluid with equation of state $p=\omega\rho$; $\omega < -1$, as well as chaplygin 
gas simultaneously then the big rip problem does not arises and the scale factor is found to be regular for all time.
\end{abstract}
\vspace{8mm}

Recent observations like CMB anisotropy, supernova and galaxy clustering strongly indicate 
that our universe is spatially 
flat and there exists an exotic cosmic fluid called dark energy with negative pressure, 
which constitutes about 70 percent of the total energy of universe. 
The dark energy is usually described by an equation of state (EoS) 
parameter $\omega\equiv\frac{p}{\rho}$, the ratio of spatially homogeneous 
dark-energy pressure p to its energy density $\rho$. 
A value $\omega<-\frac{1}{3}$ is required for cosmic acceleration. 
The simplest explanation for dark energy is cosmological constant, for which $\omega=-1$. 
The increasing evidence from observational data indicates that $\omega$ lies in a narrow strip 
around $\omega=-1$ quite likely being less than this value \cite{ref1}$-$\cite{ref3}. 
The region where EoS parameter $\omega<-1$, is typically referred to as a phantom dark energy universe. 
The existence of the region with $\omega<-1$ opens up a number of fundamental questions. 
For instance, the entropy of such universe is negative. 
The dominant energy condition (DEC) for phantom fluid is violated, as a rule. 
The phantom dominated universe end up with a finite time future singularity called big rip or 
cosmic doomsday \cite{ref4,ref5}. 
The last property attracted much attention and brought the number of speculations upto the 
explicit calculation of the rest of the life-time of our universe.\\
\par 
Soon after Caldwell \cite{ref4} proposed phantom dark energy model with cosmic doomsday of future universe, 
cosmologists started making efforts to avoid this problem using $\omega<-1$ \cite{ref6}$-$\cite{ref8}. 
In the braneworld scenario, Sahni and Shtanov has obtained well-behaved expansion for the future universe 
without big rip problem with $\omega<-1$. They have shown that acceleration is a 
transient phenomenon in the current universe and the future universe will re-enter matter 
dominated decelerated phase \cite{ref9}. It is found that general relativity (GR) based phantom 
model encounters ``sudden future singularity'' leading a divergent scale factor, energy density and pressure 
at finite time $t=t_{s}$. Thus the classical approach to phantom model exhibits big rip problem. 
For future singularity model, curvature invariant becomes very strong and energy density is very 
high near $t=t_{s}$ \cite{ref10}. So, quantum effects should be dominated for $|t=t_{s}|<$ one unit of 
time (Early universe) \cite{ref11}$-$\cite{ref13} and it is shown that the an escape from the big rip is possible 
on making quantum corrections to the energy density and pressure in Friedmann equations.\\
\par
In the framework of Robertson-Walker cosmology, Chaplygin gas (CG) is also considered as a good source of 
dark energy for having negative pressure, given as
\begin{equation}
\label{eq1}
p=-\frac{A}{\rho} 
\end{equation}
where p and $\rho$ are, respectively, pressure and energy density in a co moving reference 
frame, with $\rho>0$; A is a positive constant.\\
Moreover, it is only gas having super-symmetry generalization \cite{ref14}$-$\cite{ref16}. 
Bertolami et al \cite{ref17} have found that generalized Chaplaincy gas (CG) is better fit for 
latest Supernova data. In case of CG, equation (\ref{eq1}) is modified as
\begin{equation}
\label{eq2}
p=-\frac{A}{\rho^\frac{1}{\alpha}}
\end{equation}
where $1\leq\alpha<\infty$.\\
For $\alpha=1$, equation (\ref{eq2}) corresponds to equation (\ref{eq1}).\\
\par
Other approaches have considered dissipative effects in CG models, using the 
framework of Elkhart theory \cite{ref18}. Thai et al \cite{ref19}, have investigated a viscous GCG, 
assuming that there is a bulk viscosity in a linear borotropic fluid and GCG. It is found that 
the equation of state of GCG can cross the boundary $\omega=-1$. Also in Ref. \cite{ref20}, 
it is found that a dissipative chaplygin gas can give rise to structurally stable evaluational scenarios. It is 
interesting to note that the GCG itself can behave like a fluid with viscosity 
in the context of Eckart formalism \cite{ref18}. Fabris et al \cite{ref21}, 
have investigated an equivalence GCG and dust like fluid. Recently Cruz et al \cite{ref22}, have 
studied dissipative generalized chaplygin gas as phantom dark energy and found the cosmological 
solutions for GCG with bulk viscosity.\\ 
\par
The FRW metric for an homogeneous and isotropic flat universe is given by
\begin{equation}
\label{eq3}
ds^2=-dt^2+a(t)^2\left(dx^2+dy^2+dz^2\right)
\end{equation}
where a(t) is the scale factor and t represents the cosmic time.\\
The field equations in the presence bulk viscous stresses are
\begin{equation}
\label{eq4}
\frac{a_{4}^2}{a^2}=H^2=\frac{\rho}{3}
\end{equation}
\begin{equation}
\label{eq5}
\frac{a_{44}}{a}=-\frac{1}{6}\left(\rho+3\bar{p}\right)
\end{equation}
where $\bar{p}$ is the effective pressure given by
\begin{equation}
\label{eq6}
\bar{p}=p-3H\xi
\end{equation}
Here p, $\xi$ are the isotropic pressure and bulk viscous coefficient respectively.\\
The energy conservation equation is given by
\begin{equation}
\label{eq7}
\rho_{4}+3H\left(\rho+\bar{p}\right)=0
\end{equation}
Here, and in what follows the sub indices 4 on $a$, $\rho$ and elsewhere denote differentiation with respect to t.\\
Using equations (\ref{eq2}), (\ref{eq4}) and (\ref{eq6}), equation (\ref{eq7}) leads to
\begin{equation}
\label{eq8}
\rho_{4}+3\frac{a_{4}}{a}\left(\rho-\frac{A}{\rho^\frac{1}{\alpha}}-3H\xi\right)=0
\end{equation}
In order to obtain solution of equation (\ref{eq8}), we will assume that the viscosity has a power-law 
dependence upon the density
\begin{equation}
\label{eq9}
\xi=\xi_{0}\rho^n
\end{equation}
where $\xi_{0}$ and n are constant.\\
On using equation (\ref{eq9}) in equation (\ref{eq8}), we obtain
\begin{equation}
\label{eq10}
\frac{d\rho}{dt}+\frac{3}{a}\frac{da}{dt}\frac{\left(\rho^\frac{1+\alpha}{\alpha}-A\right)}{\rho^\frac{1}{\alpha}}=3\rho^{n+1}\xi_{0}
\end{equation}
To solve equation (\ref{eq10}), we use the transformation $\rho_{4}=f(\rho)=\rho^{n+1}$, accordingly equation (\ref{eq10}) leads to
\begin{equation}
\label{eq11}
\rho^\frac{1+\alpha}{\alpha}(t)=A+\left(\rho_{0}^\frac{1+\alpha}{\alpha}-A\right)\left[\frac{a_{0}}{a(t)}\right]^\frac{3(1+\alpha)}{\alpha(1-3\xi_{0})}
\end{equation}
where $\rho_{0}=\rho(t_{0})$ and $a_{0}=a(t_{0})$; $t_{0}$ is the present time.\\
In the present model, it is assumed that the dark energy behaves like GCG, obeying equation (\ref{eq2}) 
as well as fluid with equation of state
\begin{equation}
\label{12}
p=\omega\rho
\end{equation}
with $\omega<-1$ simultaneously.\\
From equations (\ref{eq2}) and (12), we obtain
\begin{equation}
\label{eq13}
\omega(t)=-\frac{A}{\rho(t)^\frac{1+\alpha}{\alpha}}
\end{equation}
So, evolution of equation (\ref{eq13}) at $t=t_{0}$ leads to
\begin{equation}
\label{eq14}
A=-\omega_{0}\rho_{0}^\frac{1+\alpha}{\alpha}
\end{equation}
with $\omega_{0}=\omega(t_{0})$.\\
Using equation (\ref{eq14}), equation (\ref{eq11}) leads to
\begin{equation}
\label{eq15}
\rho=\rho_{0}\left[-\omega_{0}+(1+\omega_{0})\left(\frac{a_{0}}{a(t)}\right)^\frac{3(1+\alpha)}{\alpha(1-3\xi_{0})}\right]^\frac{\alpha}{1+\alpha}
\end{equation}
In the homogeneous model of universe, a scalar field $\phi(t)$ with potential $V(\phi)$ has energy density 
\begin{equation}
\label{eq16}
\rho_{\phi}=\frac{1}{2}\phi_{4}^2 + V(\phi)
\end{equation}
and pressure
\begin{equation}
\label{eq17}
p_{\phi}= \frac{1}{2}\phi_{4}^2 - V(\phi)
\end{equation}
Equation (\ref{eq16}) and (\ref{eq17}) lead to
\begin{equation}
\label{eq18}
\phi_{4}^2=\rho_{\phi}+p_{\phi}
\end{equation}
Using equations (\ref{eq2}), (12) and (\ref{eq14}), equation (\ref{eq18}) reduces to
\begin{equation}
\label{eq19}
\phi_{4}^2=\frac{\rho^\frac{(1+\alpha)}{\alpha}+\rho_{0}^\frac{(1+\alpha)}{\alpha}\omega_{0}}
{\rho^\frac{1}{\alpha}}
\end{equation}
Equation (\ref{eq15}) and (\ref{eq19}) lead to
\begin{equation}
\label{eq20}
\phi_{4}^2=\frac{(1+\omega_{0})\rho_{0}\left(\frac{a_{0}}{a(t)}\right)^\frac{3(1+\alpha)}{\alpha(1-3\xi_{0})}}{\left[-\omega_{0}+(1+\omega_{0})\left(\frac{a_{0}}{a}\right)^\frac{3(1+\alpha)}{\alpha(1-3\xi_{0})}\right]^\frac{1}{1+\alpha}}
\end{equation}
From equation (\ref{eq20}), it is clear that for $1+\omega_{0} > 0$, $\phi_{4}^2 > 0$, 
giving positive kinetic energy and for $1+\omega_{0} < 0$, $\phi_{4}^2 < 0$, giving negative kinetic energy. $1+\omega_{0} > 0$ and 
$1 + \omega_{0} <  0$ are representing the case of quintessence and phantom fluid dominated 
universe respectively. Similar result are obtained by Hoyle and Narlikar in C -field with negative kinetic energy 
for steady state theory of universe \cite{ref23}.\\
Now, from equation (\ref{eq4}) and (\ref{eq15}), we obtain
\begin{equation}
\label{eq21}
\frac{a_{4}^2}{a^2}=\Omega_{0}H_{0}^2\left[|\omega_{0}|+(1-|\omega_{0}|)\left(\frac{a_{0}}{a(t)}\right)^
\frac{3(1+\alpha)}{\alpha(1-3\xi_{0})}\right]^\frac{\alpha}{1+\alpha}
\end{equation}
where $|\omega_{0}|=-\omega$, $H_{0}=100$hkm/s Mpc, present value of the Hubble's parameter 
and $\Omega_{0}=\frac{\rho_{0}}{\rho_{cr,0}}$ with $\rho_{cr,0}=\frac{3H_{0}^2}{8\pi G}$.\\
Equation (\ref{eq21}) may be written as
\begin{equation}
\label{eq22}
\frac{a_{4}}{a}=\sqrt{\Omega_{0}}H_{0}|\omega_{0}|^\frac{\alpha}{2(1+\alpha)}\left[1+\frac{(1-|\omega_{0}|)}{|\omega_{0}|}\left(\frac{a_{0}}{a(t)}\right)^
\frac{3(1+\alpha)}{\alpha(1-3\xi_{0})}\right]^\frac{\alpha}{2(1+\alpha)}
\end{equation}
Neglecting the higher powers of $\frac{(1-|\omega_{0}|)}{|\omega_{0}|}\left(\frac{a_{0}}{a(t)}\right)^
\frac{3(1+\alpha)}{\alpha(1-3\xi_{0})}$, equation (\ref{eq22}) leads to
\begin{equation}
\label{eq23}
\frac{a_{4}}{a}=\sqrt{\Omega_{0}}H_{0}|\omega_{0}|^\frac{\alpha}{2(1+\alpha)}\left[1+\frac{\alpha(1-|\omega_{0}|)}{2(1+\alpha)|\omega_{0}|}\left(\frac{a_{0}}{a(t)}\right)^
\frac{3(1+\alpha)}{\alpha(1-3\xi_{0})}\right]
\end{equation}
\begin{figure}
\begin{center}
\includegraphics[height=8 cm]{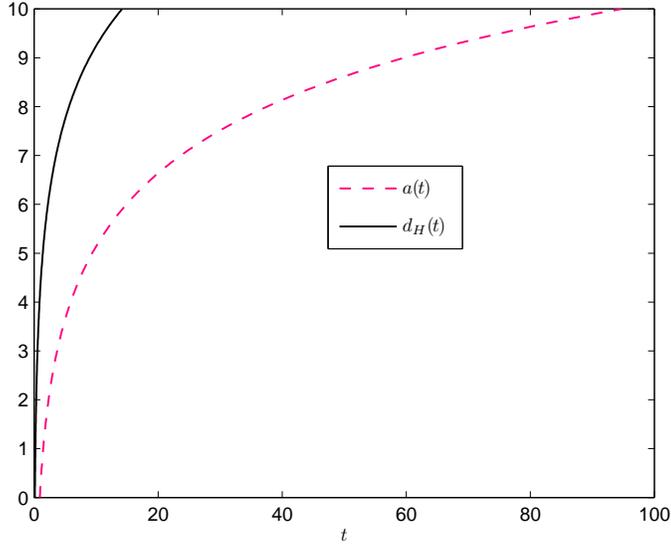}
\caption{The plot of scale factor $(a)$ and horizon distance $(d_{H})$ versus time $(t)$.}
\label{fg:anil29fig1.eps}
\end{center}
\end{figure}
\begin{figure}
\begin{center}
\includegraphics[height=8 cm]{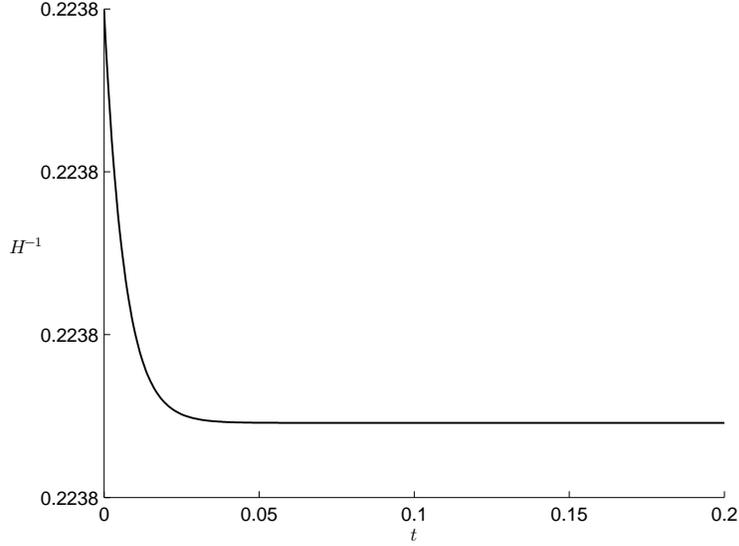}
\caption{The plot of Hubble distance $(H^{-1})$ versus time $(t)$.}
\label{fg:anil29fig2.eps}
\end{center}
\end{figure}
Integrating equation (\ref{eq23}), we obtain 
\[
 a(t)=\left(\frac{a_{0}}{\left(2(1+\alpha)|\omega_{0}|\right)}\right)^\frac{\alpha(1-3\xi_{0})}{3(1+\alpha)}\times
\]
\begin{equation}
\label{eq24}
\left[
(\alpha+2(1+\alpha)|\omega_{0}|)e^{6H_{0}|\omega_{0}|^\frac{\alpha}{2(1+\alpha)}\sqrt{\Omega_{0}}(t-t_{0})}
-\alpha(1-|\omega_{0}|)\right]^\frac{\alpha(1-3\xi_{0})}{3(1+\alpha)}
\end{equation}
From equation (\ref{eq24}), it is clear that as $t\rightarrow \infty$, $a(t)\rightarrow\infty $ which is 
supported by recent observation of Supernova Ia \cite{ref24,ref25} and WMAP \cite{ref26,ref27}. 
Therefore the present model is free from finite time future singularity. \\
Now the horizon distance is obtained as
\[
 d_{H}(t)=\frac{3(1+\alpha)a(t)}{\alpha(1-3\xi_{0})a_{0}}\left(\frac{2(1+\alpha)|\omega_{0}|}{\alpha+(\alpha+2)|\omega_{0}|}\right)^\frac{\alpha(1-3\xi_{0})}{3(1+\alpha)}\times
\]
\begin{equation}
\label{eq25}
e^{\left[6H_{0}|\omega_{0}|^\frac{\alpha}{2(1+\alpha)}\sqrt{\Omega_{0}}\frac{\alpha(1-3\xi_{0})t}{3(1+\alpha)}\right]}
\end{equation}
\textbf{Fig. 1} depicts the variation of scale factor $(a)$ and horizon distance $(d_{H})$ versus cosmic time, 
as a representative case with appropriate choice of constants and other physical parameters. 
From equation (\ref{eq24}) and (\ref{eq25}) it is clear that $d_{H}(t)>a(t)$ i. e. horizon grows more rapidly 
than scale factor which is clearly shown in \textbf{Fig. 1}. \\
In this case, the Hubble distance is given by
\begin{equation}
\label{eq26}
H^{-1}=\frac{1}{\sqrt{\Omega_{0}}H_{0}|\omega_{0}|^\frac{\alpha}{2(1+\alpha)}}\left[1-\frac{\alpha(1-|\omega_{0}|)}{2(1+\alpha)|\omega_{0}|}\left(\frac{a_{0}}{a(t)}\right)^
\frac{3(1+\alpha)}{\alpha(1-3\xi_{0})}\right]
\end{equation}
Equation (\ref{eq26}) is showing the growth of Hubble distance $(H^{-1})$ with time 
such that $H^{-1}\rightarrow \frac{1}{H_{0}\sqrt{\Omega_{0}}|\omega_{0}|^\frac{\alpha}{2(1+\alpha)}}\neq 0$ as 
$t\rightarrow\infty$. This behaviour of $H^{-1}$ is clearly depicted in \textbf{Fig. 2}. 
Thus in present case, the galaxies will not disappear when $t\rightarrow\infty$, avoiding big rip singularity. 
Therefore, one can conclude that if phantom fluid behaves like GCG and fluid with $p=\omega\rho$ simultaneously 
then the future accelerated expansion of universe will free from catastropic situation like big rip.
Equation (\ref{eq15}) may be written as
\begin{equation}
\label{eq27}
\rho=\rho_{0}\left[|\omega_{0}|+(1-|\omega_{0}|)\left(\frac{a_{0}}{a(t)}\right)^\frac{3(1+\alpha)}{\alpha(1-3\xi_{0})}\right]^\frac{\alpha}{1+\alpha}
\end{equation}
From equation (\ref{eq27}), it is clear that as 
$t\rightarrow \infty$, $\rho\rightarrow\rho_{0}|\omega_{0}|^\frac{\alpha}{1+\alpha}>\rho_{0}$ 
(since $t\rightarrow \infty$, $a(t)\rightarrow\infty $). 
Thus one can conclude that energy density increases with time, contrary to other phantom models 
having future singularity at $t=t_{s}$ \cite{ref4,ref11}. Based on Ia Supernova data, 
singh et al \cite{ref28} have estimated $\omega_{0}$ for model in the range $-2.4<\omega_{0}<-1.74$ up to 
$95$ percent confidence level. Taking this estimate as an example, with $\alpha=1$, $\rho_{\infty}$ is found in 
the range $1.31\rho_{0}<\rho_{\infty}<1.54\rho_{0}$ and with $\alpha=2$, $\rho_{\infty}$ is found in 
the range $1.44\rho_{0}<\rho_{\infty}<1.78\rho_{0}$ and so on. This does not yields much increase in 
energy density as $t\rightarrow\infty$ but if the future experiments supports large value of 
$|\omega_{0}|$ then $\rho_{\infty}$ will be high.\\

It is interesting to see here that present model is derived by Bulk viscosity  
in the context of Eckart formalism \cite{ref18}. We see that big rip problem does not arises in the 
present model. In Refs \cite{ref10}$-$\cite{ref13}, for models with future singularity escape 
from big rip, is demonstrated using quantum corrections in the field equations near $t=t_{s}$. 
However the present study deals phantom cosmology with accelerated expansion without catastrophic 
situations using classical approach. It is also seen that the model of future universe presented by 
Srivastava \cite{ref29} is a particular case  of the model presented in this letter.

\section*{Acknowledgements} 
Author would like to thanks The Institute of Mathematical Science (IMSc), Chennai, India 
for providing facility and support where part of this work was carried out. 


\end{document}